\documentclass{PoS}

\title{Hadron formation in the deconfined matter at RHIC and LHC}

\ShortTitle{Hadron formation at RHIC and LHC}

\author{\speaker{Rene Bellwied}\thanks{Thanks to my collaborators: C. Ratti, M.Barbaro (Torino) and M. Cristoforetti (Trento).}\\
        University of Houston, Physics Department\\
	617 SR1 Building, Houston, TX 77204\\
           E-mail: \email{bellwied@uh.edu}}

\abstract{We have studied the probability of two succinct particle production mechanisms, likely to occur in the transition from the deconfined medium produced in RHIC and LHC heavy ion collisions back to the hadron gas, which is abundant at lower temperatures. Evidence has been found for in-medium fragmentation of non-equilibrated partons in the deconfined medium as well as bound state formation in the quark-gluon phase itself, as described by lattice QCD . Through the comparison to PNJL model calculations we attempt to quantify an extended phase of mixed degrees of freedom in a temperature range just above the QCD transition temperature. The extracted hadron formation probabilities exhibit flavor and baryon number dependencies, which are experimentally verifiable. We confront the calculations of the non-equilibrium and equilibrium particle production in heavy ion collisions with measurements from RHIC and LHC.}

\FullConference{50th International Winter Meeting on Nuclear Physics - Bormio2012,\\
		23-27 January 2012\\
		Bormio, Italy}

\begin{document}

\section{Introduction: Formation of hadrons from a deconfined medium}

The unambiguous evidence found in RHIC heavy ion collisions for the formation of a deconfined partonic state of matter
\cite{white} enables us to investigate the mechanism of hadron formation starting from a system that mimics the state of the universe only microseconds after the Big Bang. The common interpretation has been that a finite volume deconfined state, which is thermally and chemically equilibrated, will cool down to a critical temperature at which hadronization of all particle species occurs. This temperature can be calculated on the lattice and over the decades many papers on finite temperature QCD thermodynamics have quoted temperatures in the 150-200 MeV range. With the advent of very realistic 2+1 flavor calculations and the new measurements at RHIC and the LHC a more complex hadronization pattern seems to emerge, though. Here, we will confront the formation of hadrons in fragmentation models and lattice QCD calculations with data in order to derive at a differentiated picture of particle formation in the universe. 

\subsection{ Fragmentation models}

The high momentum part of the emitted particle spectrum in relativistic heavy ion collisions is likely produced through the fragmentation of hard partons which do not equilibrated with the deconfined bulk medium. In this case the time evolution of the fragmentation process should follow the principles of vacuum fragmentation \cite{greiner}. In earlier publications \cite{mbv,bm} we have shown  that the formation time of color-neutral (pre)-hadronic states, as calculated in light-cone variables, will depend on the momentum and mass of the final state and at least some of the high momentum hadrons will be formed inside the medium \cite{mbv}. For these color-neutral bound states inside a colored medium the interaction cross section will be strongly reduced (color transparency) and measurable effects, such as a particle species dependent reduction in the nuclear suppression factor at intermediate transverse momentum, is possible \cite{bm}. Evidence for such a behavior (originally referred to as baryon puzzle) has been found at RHIC and more recently at the LHC \cite{appels}. 

\subsection{Lattice QCD predictions}

\begin{figure}
\begin{minipage}{.48\textwidth}
\parbox{6cm}{
\scalebox{.6}{
\includegraphics{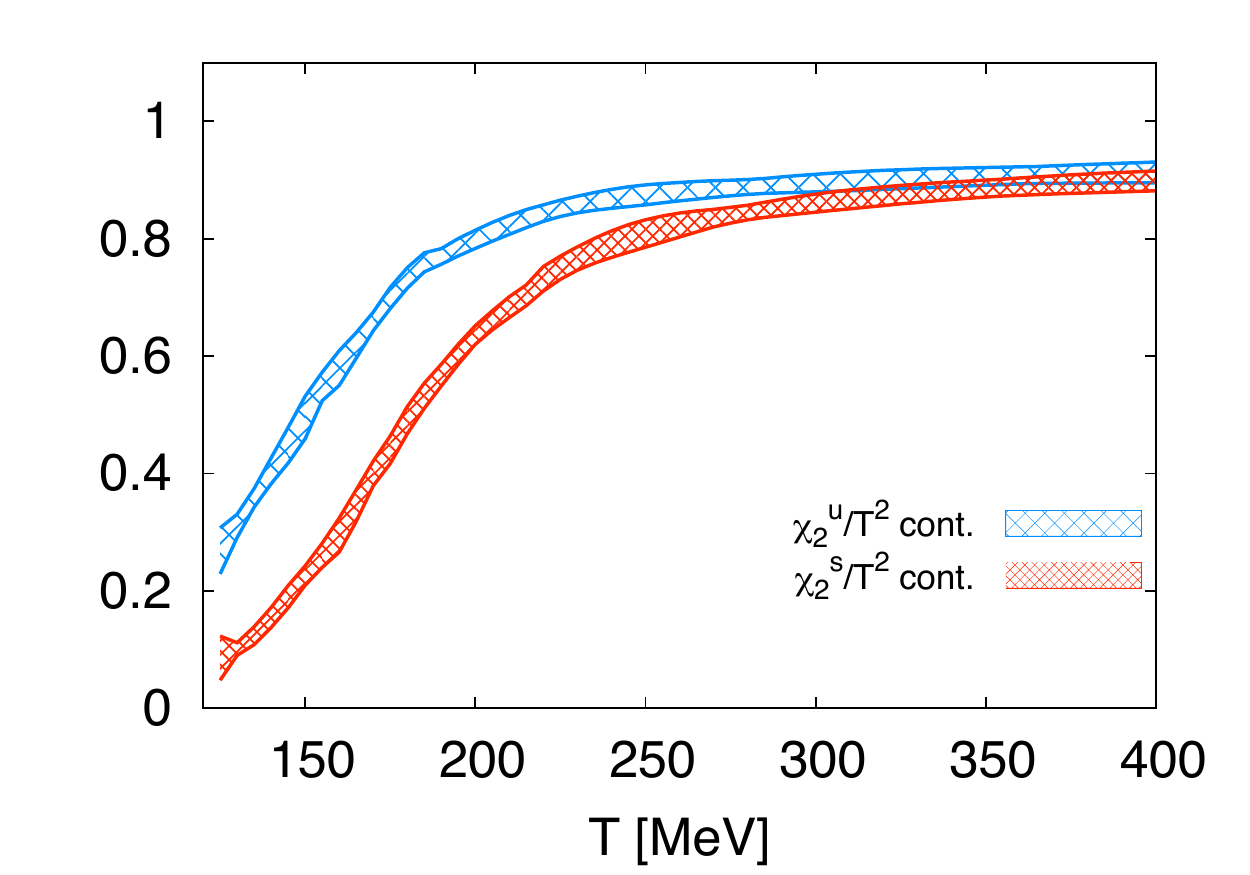}\\}}
\end{minipage}
\begin{minipage}{.48\textwidth}
\parbox{6cm}{
\scalebox{.6}{
\includegraphics{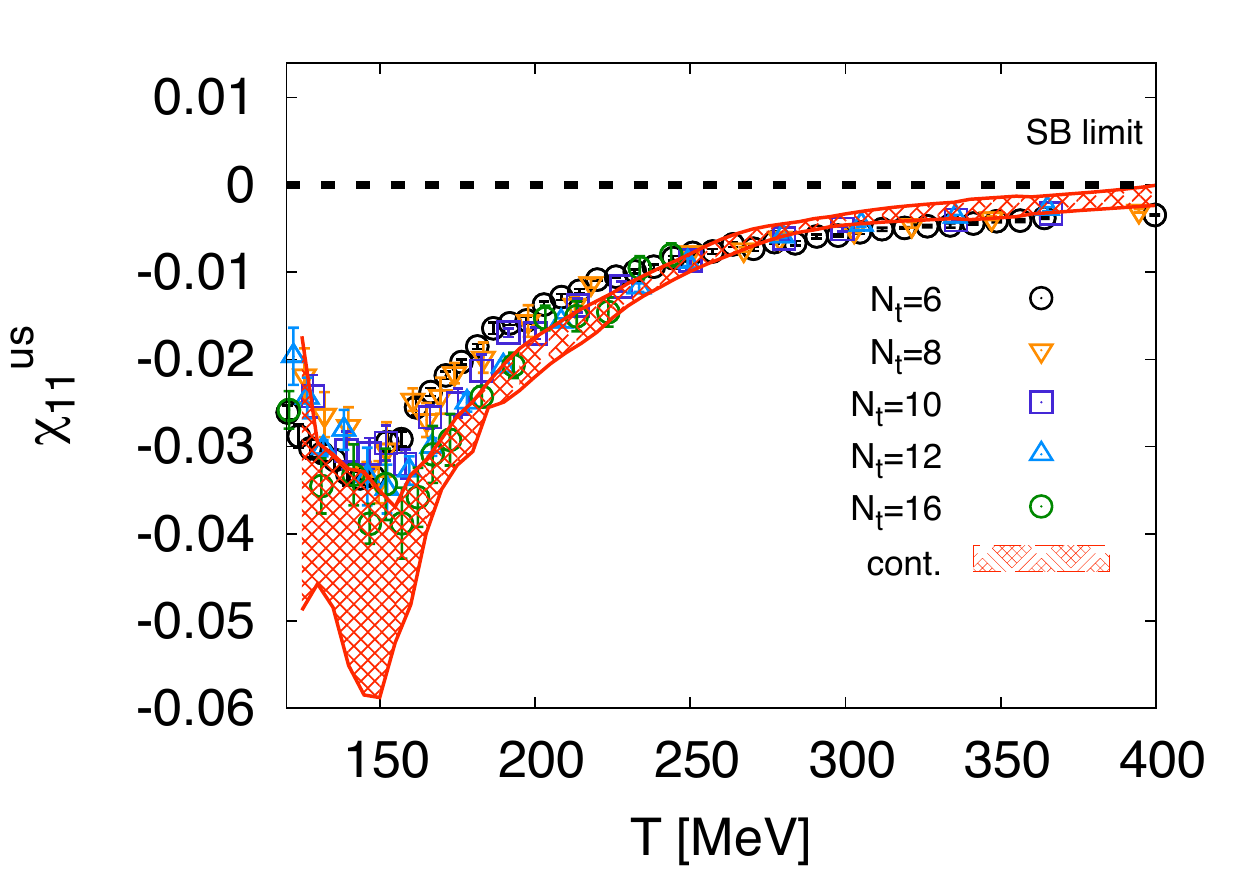}\\}}
\end{minipage}
\caption{Comparison between the lattice result continuum limits for light and strange quark number susceptibilities, obtained with the stout action at physical quark masses (left plot)  and non-diagonal $u-s$ susceptibilities as a function of lattice spacing (right plot) in the Wuppertal-Budapest collaboration approach \cite{wb}. }
\label{fig1}
\end{figure}

Over the past three decades the continuous improvement of lattice QCD calculations due smaller lattice spacings, better gauge actions and the use of physical quark and hadron masses, including the pion, have led to a significant revision
of the conclusions regarding the nature of the QCD phase transition. The most recent lattice QCD calculations, with lattice spacings that approach the continuum limit,  have revealed that the transition from hadronic to free
partonic degrees of freedom at zero baryo-chemical potential is merely an analytic cross-over rather than the anticipated second order phase transition  \cite{Aoki:2006we}. The main quantities that are used to determine the transition temperature are the Polyakov loop, energy density and quark number susceptibilities for the deconfinement phase transition, and the quark condensates for the chiral phase transition. The smooth behavior of all these QCD observables as functions of the temperature  leads to interesting cross-over phenomena \cite{Borsanyi:2010bp,plumari}. Both deconfinement, as expressed through the renormalized Polyakov loop or quark number susceptibilities, and chiral symmetry restoration, shown in the chiral condensates, experience an extended transition region, up to 2 $T_{c}$, before reaching the fully deconfined and chirally symmetric state. At any given temperature in this cross-over range these parameters could thus be interpreted as signalling a mixed phase of degrees of freedom where bound states or chirally broken states will co-exist with free quarks and gluons according to the relative values of the Polyakov loop or the chiral condensates. This transition region was called ``semi-QGP" in Ref. \cite{Hidaka:2008dr}, as opposed to the ``full-QGP" at large temperatures, where the Polyakov loop is close to one and flat.
Furthermore, the latest lattice results signal a flavor dependence of quark number susceptibilities even in the light quark sector. The rise of the strange quark susceptibility with temperature is slower and takes place at larger temperatures  compared to the $u$ case, as shown in  Fig.\ref{fig1}(left). This feature was less pronounced in previous lattice results \cite{eos}, since in that case $m_s/m_{u,d}$ = 10, whereas the new results use a physical quark mass ratio ($m_s/m_{u,d}$ = 28.15).

This flavor difference between quark number susceptibilities in the light sector likely indicates that strange quarks experience deconfinement at slightly larger temperatures, compared to light quarks, thus implying a survival of strangeness-carrying hadrons in the QGP immediately above $T_c$.

Non-diagonal $u-s$ susceptibilities (shown in Fig.\ref{fig1}( right)) exhibit a pronounced peak in the vicinity of the phase transition, and remain finite for relatively large temperatures above $T_c$. Although it was shown in Ref. \cite{Blaizot:2001vr} that correlations between different flavors are nonzero in perturbative QCD at large temperatures due to the presence of flavor-mixing diagrams, the lattice data exhibit a strong enhancement of these correlations in the vicinity of $T_c$, which survives up to relatively large temperatures above the transition \cite{wb,Borsanyi:} and which cannot be accounted for by the perturbative QCD contribution alone. Taking into account this behavior, one could again come to the conclusion that in the region 1-2 $T_c$ the probability of forming color neutral bound states is quantifiable even in the case of a fully equilibrated system of quarks and gluons as simulated through lattice QCD.

\section{Comparison to PNJL calculations and hadron resonance gas models}

In order to draw conclusions on the presence of bound states in the QGP, we compare lattice results for non-diagonal quark correlators to calculations performed with a PNJL model \cite{ours}. The u-s correlators on the lattice should include,  besides the quark contributions, also contributions from hadronic bound states, in  particular the Kaon. The PNJL model on the other hand is tuned in a way that all fluctuations (of the Polyakov loop and the chiral and quark condensates) are taken into account but no hadronic contribution is calculated. Thus we expect that any difference between lattice and PNJL results will quantify the existence of hadronic bound states above the QCD transition temperature, as determined by the inflection point in the non-diagonal correlator curve as a function of temperature.  

\begin{figure}
\begin{minipage}{.48\textwidth}
\hspace{-.8cm}
\parbox{6cm}{
\scalebox{.66}{
\includegraphics{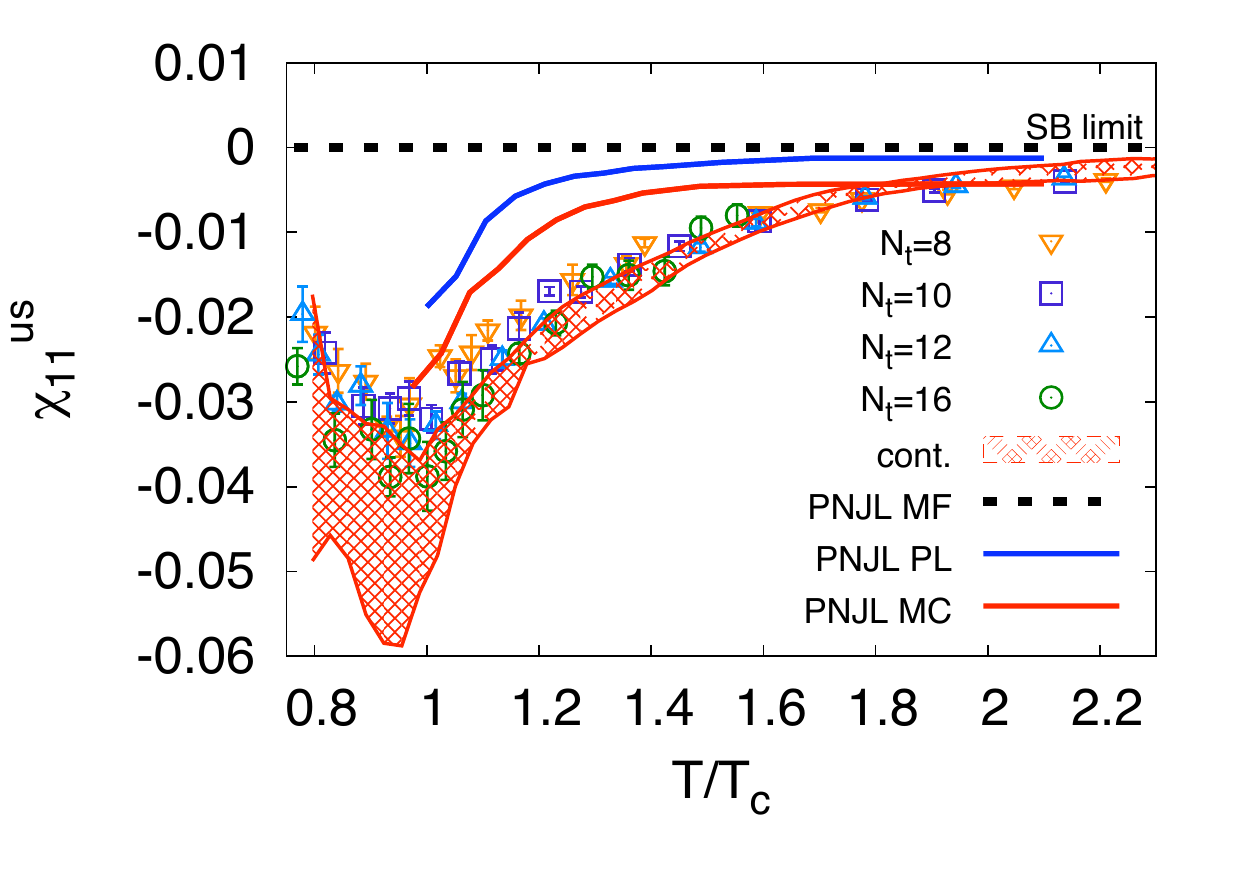}\\}}
\end{minipage}
\begin{minipage}{.48\textwidth}
\parbox{6cm}{
\scalebox{.64}{
\includegraphics{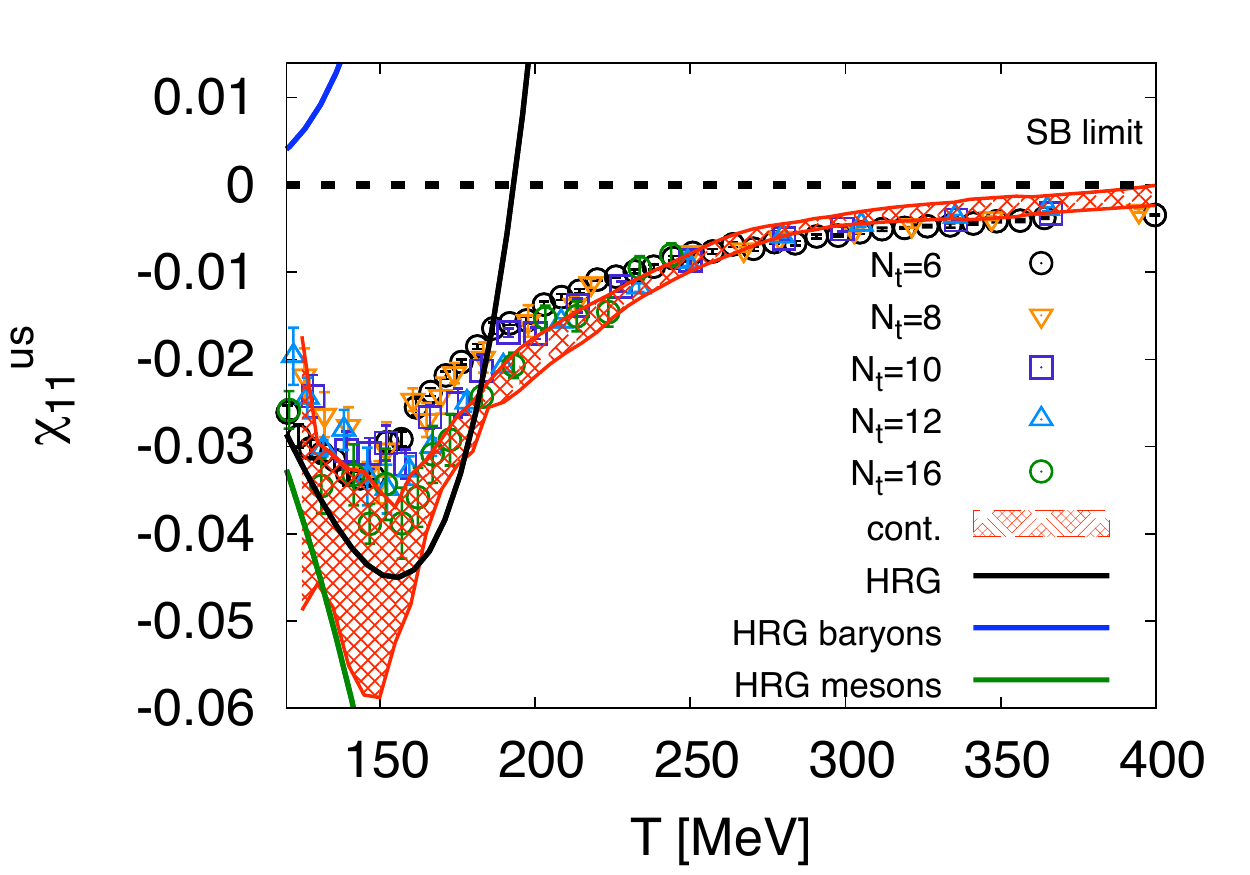}\\}}
\end{minipage}
\caption{Left: Comparison between the lattice results for the $u-s$ correlator as a function of $T/T_c$ \cite{wb}, and the PNJL model results. The mean field PNJL result is zero for all temperatures, as expected (dashed curve). The blue curve corresponds to the PNJL model result when only the Polyakov loop fluctuations are taken into account. The red curve is the full PNJL model prediction, with fluctuations of all fields taken into account. Notice that the red curve will fall on the blue curve in the infinite volume limit. Right: Comparison between the lattice results for the $u-s$ correlator and Hadron Resonance Gas model results. The different curves show the contributions of baryons (blue), mesons (green) and the total (black) from the HRG model \cite{hrg}.}
\label{fig2}
\end{figure}

At the mean field level, the PNJL model has no correlations between the different quark flavors, therefore the corresponding $u-s$ correlator stays flat and equal to zero over the full temperature range. In order to properly estimate
all possible contributions to this observable from colored degrees of freedom and mesonic zero modes, we need to go beyond mean field and take fluctuations of all fields (Polyakov loop, chiral condensates, pion and kaon condensates) into account. 
The importance of including fluctuations in the model is shown in Fig.\ref{fig2}(left), where we compare the chiral condensate from lattice QCD (from Ref. \cite{wb}), the PNJL result at the mean-field level, and the PNJL result in which fluctuations of all fields are included. As it is evident, the inclusion of fluctuations makes the curve much smoother and brings it closer to the lattice data (a similar effect was observed in Ref. \cite{Skokov:2010wb}).

Still, the correlator values, in comparison to PNJL model calculations beyond mean-field in the temperature regime between 1-2 $T_c$, indicate that at least part of the mixed phase resides in color-neutral bound states.

Furthermore, the peculiar shape of the non-diagonal correlator near $T_c$ (sharp dip and subsequent slow rise towards zero) can be interpreted when comparing it to the separate contributions of mesonic and baryonic states in a hadron resonance gas (HRG) calculation \cite{hrg}. Mesonic states in the HRG model exhibit a negative correlation, whereas baryonic states yield a positive value (see Fig.\ref{fig2}(right)). The dip and slow rise thus is likely caused by enhanced baryonic state formation (or survival) at higher temperatures. Still, the lattice data never exceed zero before full deconfinement is reached, which means that baryonic states never dominate the hadron formation. 

\begin{figure}
\begin{center}
\parbox{6cm}{
\scalebox{.64}{
\includegraphics{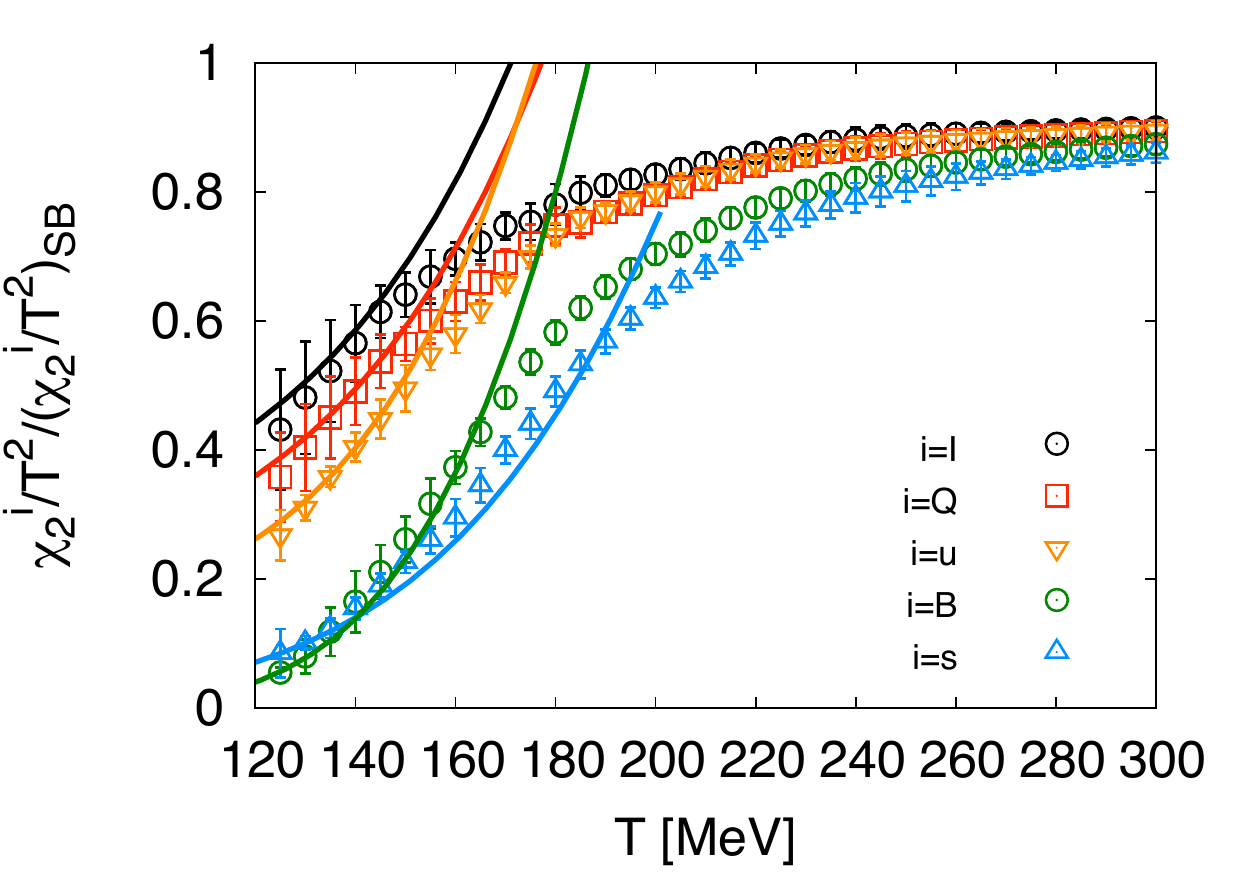}\\}}
\end{center}
\caption{Summary of lattice QCD continuum extrapolations of all relevant quark number susceptibilities (isospin, charge, baryon number, light quark, strange quark) in the latest calculations by the Wuppertal-Budapest collaboration \cite{wb}. The plot suggest a higher transition temperature for strange quark and baryonic states. }
\label{fig3}
\end{figure}

Taking into account the flavor and baryon number dependencies of all light quark susceptibilities, as shown in Fig.\ref{fig3}, one recognizes two distinct quantum number groupings and thus can deduce a scenario where strange quark bound states are formed (or survive) 
at higher $T$ in the deconfined medium than light quark bound states. Earlier lattice calculations only exhibited this effect in the comparison between light and heavy quarks, but the recent improvement in lattice accuracy indicates effects already at the
strange quark level, which leads to specific experimentally verifiable predictions, such as an enhanced survival probability of strange over non-strange resonances near, but above, $T_c$.

\section{Experimental evidence}

The possibility of a flavor and baryon number dependent hadronization pattern as a function of temperature  has recently received indirect confirmation through two key measurements by the ALICE collaboration.

In the high momentum range the apparent difference in the energy loss of baryons and mesons as measured through the nuclear suppression factors of Kaons and Lambdas \cite{appels} can be attributed to differing flavor and mass dependent formation times of the non-equilibrated hadrons in the deconfined medium and subsequent color transparency \cite{bm}. One should note though that alternative explanations based on direct baryon production \cite{brodsky} or recombination \cite{fries} are also possible.

Even more convincing might be the evidence for flavor dependent emission from the equilibrated medium as measured through particle identified yield ratios. The ALICE particle ratios in the light and strange quark sector \cite{alice-ratios} cannot be reconciled with a single temperature statistical hadronization model \cite{andronic} but require different temperatures for light quark and strange quark ratios. The extracted temperature difference agrees well with the shift in the inflection points of the light and strange quark correlators as shown in Fig.1(left). 

\section{Conclusions}

In summary, a comparison of PNJL to lattice QCD calculations yields ample evidence, that a phase of mixed degrees of freedom exists for a particular temperature range above the QCD transition temperature. The PNJL model calculations and the mapping of the flavor and baryon number dependence in lattice QCD calculations, show that it is likely that in this equilibrated phase the relative abundance of strange and baryonic bound states is larger the higher the temperature. In other words, strange hadrons and non-strange baryons form earlier in the mixed phase than their non-strange and mesonic partners. Still, mesonic bound states dominate over the entire temperature range.  These dependencies are now experimentally verifiable through detailed particle identified measurements of the properties of light-, strange- and charm-quark hadrons and resonances near the phase boundary. First evidence is given by differences in the particle identified nuclear suppression factors, as well as the strange particle to pion ratios compared to the proton to pion ratios in ALICE.

\end{document}